\documentclass[prd,twocolumn,preprintnumbers,superscriptaddress,nofootinbib,floatfix]{revtex4}
\usepackage[a4paper, hdivide={1.91cm,,1.165cm}, vdivide={1.83cm,,3.6cm}]{geometry}

\usepackage{amstext,amssymb}
\usepackage{amsmath}
\usepackage{graphicx}
\usepackage[hyperfootnotes=false]{hyperref}
\usepackage{xspace}
\usepackage{color}
\usepackage{units}
\usepackage{slashed} 

\begin{document}

\title{Singlet fermion Dark Matter within Left-Right Model}
\author{Sudhanwa Patra}
\email{sudha.astro@gmail.com}
\affiliation{Center of Excellence in Theoretical and Mathematical Sciences, \\
 \hspace*{-0.0cm} Siksha 'O' Anusandhan University, Bhubaneswar-751030, India}
 \author{Soumya Rao}
\email{soumya.rao@adelaide.edu.au}
\affiliation{ARC Centre of Excellence for Particle Physics at the Terascale, Department of Physics, \\
University of Adelaide, Adelaide, SA 5005, Australia.}
\begin{abstract}

We discuss singlet fermion dark matter within a left-right symmetric model promoting
baryon and lepton numbers as separate gauge symmetries. We add a simple Dirac fermionic
dark matter singlet under $SU(2)_{L,R}$ with nonzero and equal baryon and lepton number
which ensures electric charge neutrality.  Such a dark matter candidate interacts with SM particles
through the extra $Z_{B,\ell}$ gauge bosons.  This can give rise to a dark matter particle
of a few hundred GeV that couples to $\sim$ TeV scale gauge bosons to give the correct
relic density.  This model thus accommodates TeV scale $Z_{B,\ell}$ gauge bosons and other
low scale BSM particles, which can be easily probed at LHC.

\end{abstract}
\maketitle 
\section{Introduction} 

Standard Model (SM) has proven to be a highly successful theory in the history of particle
physics accounting for forces and interactions between known fundamental particles up to
current accelerator energy.  However, one of the unexplained problems in SM is the
existence of  Dark matter (DM).  There are many possible candidates for DM with the Weakly
Interacting Massive Particle (WIMP) scenario being one of most well studied.  The
left-right symmetric model
(LRSM)\cite{Mohapatra:1974gc,Pati:1974yy,Senjanovic:1975rk,Senjanovic:1978ev,
Mohapatra:1979ia,Mohapatra:1980yp} provides a framework for incorporating potential DM
candidates in a beyond SM (BSM) scenario with the introduction of additional
multiplets\cite{Heeck:2015qra}.  We aim here to explore the possibility of studying DM
phenomenology in a special class of LRSM where DM mass can be of the order of a few
hundred GeV. 



The range of DM mass in LRSM can be from $\sim$ keV to $\sim$ TeV.  It is noted
in refs.  \cite{Bezrukov:2009th,Nemevsek:2012cd} that the keV scale right
handed neutrino can be a long-lived warm DM
candidate\footnote{Ref.\cite{Barry:2014ika} studies the interesting signatures
of these keV right handed neutrinos in neutrino mass searches.}.  However, it
is known that such a DM candidate is overabundant and needs a delicate
production mechanism in early universe which is not very natural.  Very
recently a novel approach \cite{Heeck:2015qra} was taken to introduce stable
TeV scale DM where stability of the DM is ensured either by the remnant
discrete symmetry or accidentally via high dimensionality of DM multiplets
forbidding tree-level decays.  This approach involved DM candidates as
fermionic triplets or quintuplets or scalar doublets and seven-plets.  The
detailed phenomenology of these DM candidates has been studied very recently in
\cite{Garcia-Cely:2015quu}.  

The higher $SU(2)_{L,R}$ dimensionality of stable dark matter in these models
means that the constraints from PLANCK-WMAP and indirect detection push the DM
mass beyond the reach of LHC.  On the other hand if the DM particle is a
singlet, it will be able to satisfy relic density bounds from
PLANCK\cite{Ade:2015xua} data as well as indirect detection constraints
even at lower DM masses of a few hundred GeV.  Motivated by the phenomenology
of singlet DM we consider a simple LRSM where baryon and lepton numbers are
separate local gauge symmetries \footnote{Gauge theories of baryons and leptons
has been discussed in ref.\cite{Duerr:2013dza}.}.  In this letter we study the
framework in which the DM is a LR singlet with equal B and L charges thus
ensuring its electric charge neutrality.

\section{Conventional Left-Right Models}

The prime goal here is to discuss conventional left-right symmetric model and demonstrate
that why it is difficult to accommodate singlet dark matter. The basic gauge group of
conventional left-right symmetric model \cite{Mohapatra:1974gc, Pati:1974yy,
Senjanovic:1975rk,Senjanovic:1978ev, Mohapatra:1979ia,Mohapatra:1980yp} is given by
$$\mathcal{G}_{L,R}\equiv SU(2)_{L} \otimes SU(2)_{R} \otimes U(1)_{B-L},$$ where $B-L$ is
the difference between baryon and lepton number. The electric charge is related to the 3rd
component of isospin for $SU(2)_{L,R}$ gauge group and $B-L$ charge as  
\begin{align}
Q=T_{3L}+T_{3R}+\frac{B-L}{2}
\label{eq:charge}
\end{align}
The usual  quarks and leptons belong to following representations
\begin{eqnarray}
q_{L}=\begin{pmatrix}u_{L}\\
d_{L}\end{pmatrix}\equiv[2,1,{\frac{1}{3}}] & , & q_{R}=\begin{pmatrix}u_{R}\\
d_{R}\end{pmatrix}\equiv[1,2,{\frac{1}{3}}]\,,\nonumber \\
\ell_{L}=\begin{pmatrix}\nu_{L}\\
e_{L}\end{pmatrix}\equiv[2,1,-1] & , & \ell_{R}=\begin{pmatrix}\nu_{R}\\
e_{R}\end{pmatrix}\equiv[1,2,-1] \nonumber
\end{eqnarray}

The spontaneous breaking of left-right symmetric models is implemented with i) Scalar
bidoublet $\Phi(2,2,0)$ plus doublets $H_L(2,1,-1)$+$H_R(1,2,-1)$, ii) Scalar bidoublet
$\Phi(2,2,0)$ plus scalar triplets $\Delta_L(3,1,-2)$+$\Delta_R(1,3,-2)$.  However, if we
introduce a dark matter particle singlet under $SU(2)_{L,R}$ and charged under
$U(1)_{B-L}$-- looking at electric charge formula--it is found that there is no way to
find a electrically neutral stable dark matter candidate. This gives us strong motivation
to consider alternative class of left-right symmetric model which can accommodate singlet
dark matter discussed in following section.

\section{The Present Model Framework}

We go beyond the conventional left-right symmetric models and construct a very simple
model of left-right theory accommodating singlet dark matter.  The basic gauge group of
left-right theory where individual baryon and lepton number promoted as local gauge
symmetry is given by\cite{He:1989mi} $$\mathcal{G}^{BL}_{L,R}\equiv SU(2)_L \times SU(2)_R
\times U(1)_{B} \times U(1)_{L}$$--omitting the $SU(3)_C$ structure for simplicity. The
standard quarks and leptons transforming under this new class of left-right symmetric gauge
group are
\begin{eqnarray}
q_{L}=\begin{pmatrix}u_{L}\\
d_{L}\end{pmatrix}\equiv[2,1,{\frac{1}{3}}, 0] & , & q_{R}=\begin{pmatrix}u_{R}\\
d_{R}\end{pmatrix}\equiv[1,2,{\frac{1}{3}},0]\,,\nonumber \\
\ell_{L}=\begin{pmatrix}\nu_{L}\\
e_{L}\end{pmatrix}\equiv[2,1,0, 1] & , & \ell_{R}=\begin{pmatrix}\nu_{R}\\
e_{R}\end{pmatrix}\equiv[1,2,0,1] \nonumber
\end{eqnarray}
It is known that additional $U(1)$ gauge groups introduces extra gauge anomalies to the
theory which needs to be canceled. These gauge anomalies in case of extra $U(1)_{B}$ and
$U(1)_L$ gauge groups are
\begin{align}
&\mathcal{A}\left[ SU(2)^2_L\times U(1)_B\right]=3/2\, , \nonumber \\
&\mathcal{A}\left[ SU(2)^2_R\times U(1)_B\right]=-3/2\, , \nonumber \\
&\mathcal{A}\left[ SU(2)^2_L\times U(1)_L\right]=3/2\, , \nonumber \\
&\mathcal{A}\left[ SU(2)^2_R\times U(1)_L\right]=-3/2\, , \nonumber 
\end{align}

\begin{table}[ht]
\begin{center}
\begin{tabular}{|c|c|c|c||c|c|}
\hline
 & Field                        & $ SU(2)_L$ & $SU(2)_R$ & $U(1)_B$ & $U(1)_L$ \\
\hline
\hline
Fermions&$q_L$ &  2         & 1         & 1/3      & 0    \\[1mm]
   & $q_R$     &  1         & 2         & 1/3      & 0    \\[1mm]
   & $\ell_L$  &  2         & 1         & 0        & 1    \\[1mm]
   & $\ell_R$  &  1         & 2         & 0        & 1    \\[1mm]
   \hline
   &$\Sigma_L$     &  3         & 1         & -3/4     & -3/4    \\[1mm]
   &$\Sigma_R$     &  1         & 3          & -3/4    & -3/4    \\[1mm]
   &$(\chi_L,\chi_R)$          &   1        & 1           & $n_B$& $n_L$       \\[1mm]
\hline
Scalars&$\Phi$                    &  2         & 2          & 0     &  0    \\[1mm]
   &$\Delta_L$                    &  3         & 1          & 0     &  -2    \\[1mm]
   &$\Delta_R$                    &  1         & 3          & 0     &  -2    \\[1mm]
   &$S_{BL}$                         &  1         & 1          & 3/2   &  3/2    \\[1mm]
\hline
\hline
\end{tabular}
\end{center}
\caption{Particle content of LR model where Baryon and Lepton number individually gauged. The baryon $n_B$ and 
lepton charges $n_L$ are same in order to ensure electromagnetic charge neutrality of the Dirac fermion 
$\chi=\chi_L \oplus \chi_R$.}
\label{tab:LRBL}
\end{table}

along with other vanishing gauge anomalies \hspace*{-0.2cm}$\mathcal{A}\left[
U(1)^2_{B(L)}\times U(1)_{L(B)}\right]$, $\mathcal{A}\left[ \mbox{gravity}\times
U(1)_{L(B)}\right]$ and $\mathcal{A}\left[ U(1)^3_{L(B)}\right]$. Moreover it is already
pointed out in ref.\cite{He:1989mi} that there are various ways to construct a anomaly
free left-right symmetric model with gauging baryon and lepton numbers by adding extra
pair of lepto-quarks \footnote{A viable model of left right theories promoting B and L as
gauge symmetries and its connection to neutrino mass via type III seesaw has been studied
in \cite{Duerr:2013opa}.} transforming under the gauge group $SU(3)_C \times SU(2)_L
\times SU(2)_R \times U(1)_B \times U(1)_L$ as follows:

\begin{eqnarray}
&\hspace*{-0.5cm}{\bf A.}& \Sigma_L \sim (1,3,1,-\frac{3}{4},-\frac{3}{4}),\Sigma_R \sim (1,1,3,-\frac{3}{4},-\frac{3}{4}),\nonumber \\
&\hspace*{-0.5cm}{\bf B.}& \Psi_L \sim (1,2,1,-n,-n), \quad  \Psi_R \sim (1,1,2,-n,-n),\nonumber \\
&\hspace*{-0.5cm}{\bf C.}& \Psi_L \sim (3,2,1,-\frac{n}{3},-\frac{n}{3}), \Psi_R \sim (3,1,2,-\frac{n}{3},-\frac{n}{3}),\nonumber \\
&\hspace*{-0.5cm}{\bf D.}& \Psi_L \sim (N,2,1,-\frac{n}{N},-\frac{n}{N}),   \Psi_R \sim (N,1,2,-\frac{n}{N},-\frac{n}{N}),\nonumber 
\end{eqnarray}

where $n$ is the number of fermion generation and $N$ is the dimension of these
lepto-quarks under $SU(3)_C$ gauge group. 

We propose an anomaly free left-right symmetric model promoting baryon and lepton numbers
as separate gauge symmetries. In addition to the usual quarks, $q_L(2,1,1/3,0)$,
$q_R(1,2,1/3,0)$ and leptons, $\ell_L(2,1,,0,1)$ and $\ell_R(1,2,0,1)$ we include extra
lepto-quarks $\Sigma_L(3,1,-3/4,-3/4)$ and $\Sigma_R(1,3,-3/4,-3/4)$ for anomaly
cancellation. The role of these lepto-baryons $\Sigma_{L,R}$ is to generate neutrino mass
via type-III seesaw mechanism which has been pointed out in \cite{Duerr:2013opa}.  Instead
in this letter we intend to discuss stable cold dark matter which can not only satisfy
relic density consistent with PLANCK data and indirect detection constraints but can also
give novel collider possibility. This can be easily incorporated by introducing a dark
matter candidate which is a fermionic singlet under $SU(2)_{L,R}$ as $\chi(1,1,n_B,
n_\ell)$ with the electrically charge neutral condition of DM impose equal value of $n_B$
and $n_\ell$.  This singlet dark matter can be applicable to other scenarios also.  The
Higgs sector of the model consists of a bidoublet $\Phi \equiv \left( 2_L, 2_R, 0_{B},
0_{\ell}\right)$ and two triplet scalars, $\Delta_L \equiv  \left( 3_L, 1_R, 0_{B},
2_{\ell}\right)$ and $\Delta_R  \left(1_L, 3_R, 0_{B}, 2_{\ell}\right)$ \begin{eqnarray}
	&&\Delta_{L,R} \equiv \begin{pmatrix} \delta_{L,R}^+/\sqrt{2} & \delta_{L,R}^{++}
		\\ \delta_{L,R}^0 & -\delta_{L,R}^+/\sqrt{2} \end{pmatrix}\,, \Phi \equiv
	\begin{pmatrix} \phi_1^0 & \phi_2^+ \\ \phi_1^- & \phi_2^0 \end{pmatrix}\,.
	\nonumber \end{eqnarray} along with singlet scalar $S_{BL}(1_L, 1_R,
3/2_B,3/2_\ell)$. The spectrum is presented in Table\,\ref{tab:LRBL}.

At the first stage the symmetry is broken down to LR gauge group via $S_{BL}$ by breaking
baryon and lepton number symmetries while preserving $B-L$. This singlet scalar also gives
a Majorana mass term to fermion triplet once it takes a non-zero VEV. The second stage of
symmetry breaking can be done via Higgs triplets. These triplets are needed for giving
Majorana masses to neutrinos via respective VEVs of these Higgs triplets $\langle
\Delta_{L,R}\rangle = v_{L,R}$.  Lastly, the electroweak symmetry is broken by standard
Higgs doublet belonging to a bi-doublet $\Phi$ with VEV $\langle \Phi \rangle
=\mbox{diag}(v_1, v_2)$. The hierarchy between different VEVs is $v^2_L \ll v^2=v^2_1 +
v^2_2 \ll v^2_R, v_{BL}$.

We present here the Lagrangian for the present model as
\begin{eqnarray}
\mathcal{L}^{\rm BL}_{\rm LR}&=& \mathcal{L}^{\rm }_{\rm scalar} + \mathcal{L}^{\rm gauge}_{\rm Kin.} 
      + \mathcal{L}^{\rm fermion}_{\rm Kin.} + \mathcal{L}_{\rm Yuk}\,.
\end{eqnarray}
The different parts of the Lagrangian can then be written as follows.  Firstly, the
scalar Lagrangian is written as
\begin{eqnarray}
\mathcal{L}^{\rm }_{\rm scalar} &=&
      \mbox{Tr}\big[\left(\mathcal{D}_\mu \Phi\right)^\dagger \left(\mathcal{D}^\mu \Phi\right) \big] 
      + \mbox{Tr}\big[\left(\mathcal{D}_\mu \Delta_L\right)^\dagger \left(\mathcal{D}^\mu \Delta_L\right) \big]  \nonumber \\
      &+& \mbox{Tr}\big[\left(\mathcal{D}_\mu \Delta_R\right)^\dagger \left(\mathcal{D}^\mu \Delta_R\right) \big] 
      + \left(\mathcal{D}_\mu S_{BL}\right)^\dagger \left(\mathcal{D}^\mu S_{BL}\right)  \nonumber \\
      &-& \mathcal{V}(\Phi, \Delta_L, \Delta_R, S_{BL}) 
\end{eqnarray} 
where $\mathcal{V}(\Phi, \Delta_L, \Delta_R, S_{BL})$ is the scalar potential and $\mathcal{D}_\mu$ is the 
covariant derivative for respective scalars. 

With the additional $U(1)_{B,L}$ gauge groups, the Lagrangian for kinetic terms for gauge
bosons is given by
\begin{eqnarray} 
\mathcal{L}^{\rm gauge}_{\rm Kin.}&=&
     -\frac{1}{4}W_{\mu\nu L}.W^{\mu\nu L}-\frac{1}{4}W_{\mu\nu R}.W^{\mu\nu R}  \nonumber \\
     &-&\frac{1}{4} Z^B_{\mu\nu}Z^{\mu\nu,B} -\frac{1}{4}Z^\ell_{\mu\nu}Z^{\mu\nu,\ell}
\end{eqnarray}
while that for fermions by
\begin{eqnarray} 
\mathcal{L}^{\rm fermion}_{\rm Kin.}&=&
       i  \overline{q_{L}}\gamma^{\mu} \mathcal{D}_\mu q_{L}
      +i  \overline{q_{R}}\gamma^{\mu} \mathcal{D}_\mu q_{R}  
      +i  \overline{\ell_{L}}\gamma^{\mu} \mathcal{D}_\mu \ell_{L} \nonumber \\
      &+&i  \overline{\ell_{R}}\gamma^{\mu} \mathcal{D}_\mu \ell_{R}
    +i\overline{\Sigma_{L}}\gamma^{\mu} \mathcal{D}_\mu \Sigma_{L}
      +i  \overline{\Sigma_{R}}\gamma^{\mu} \mathcal{D}_\mu \Sigma_{R} \nonumber \\
\end{eqnarray}
Now we can define the respective covariant derivatives, in general, as
\begin{eqnarray} 
\mathcal{D}^\Psi_\mu \Psi&=&\bigg[\partial_{\mu}-i\,g_L \tau^a W^a_{\mu L} -i\,g_R \tau^a W^a_{\mu R} \nonumber \\
                  &&-i\,g_B B^\Psi Z^B_{\mu}-i\,g_\ell \ell^\Psi Z^\ell_{\mu} \bigg] \Psi
\end{eqnarray}

The Yukawa structure of the present framework is
\begin{eqnarray}
\mathcal{L}_{\rm Yuk}&=&
     Y_q\, \overline{q_{L}} \Phi q_{R} + \widetilde{Y_q} \overline{q_{L}}  \widetilde{\Phi}\, q_{R} \nonumber \\
     &&+ Y_\ell\, \overline{\ell_{L}} \Phi \ell_{R} + \widetilde{Y_\ell} \overline{\ell_{L}} \widetilde{\Phi} \ell_{R}\nonumber \\
     &&+ \big[{f_L}_{}\overline{(\ell_{L})^c} (i\tau_2) \Delta_L \ell_{L}
	+{f_R}_{} \overline{(\ell_{R})^c} (i\tau_2) \Delta_R \ell_{R} \big]\nonumber \\
	&&+\lambda_\Sigma \left(\Sigma^T_L C \Sigma_L +\Sigma^T_R C \Sigma_R \right)\,S_{BL}\nonumber \\
	&&+M_\chi \overline{\chi} \chi
                       +\,  \text{h.c.}
\end{eqnarray}
\subsection{Gauge Boson Mass}

In the present framework the mass matrix of the weak gauge bosons
$(W_{\mu}^{3L}, W_{\mu}^{3R}, Z_\mu^{\ell}, Z_\mu^{B})$, is given by
\begin{eqnarray}
\hspace*{-0.5cm}
\left(\begin{array}{cccc}
\frac{1}{4} g^2_L (v^2)        &-\frac{1}{4}  g_L g_R (v^2)       & 0 &  0 \\
-\frac{1}{4} g_L g_R (v^2)    & \frac{1}{4} g^2_R (v^2 + 4 v_R^2 )      & 2 g_R g_{\ell} v_R^2 & 0 \\
0  & 2 g_R g_\ell v_R^2  & \frac{9}{4}{g_{\ell}}^2 v_{BL}^2 + 4 g_\ell^2 v_R^2 &  \frac{9}{4} g_B g_\ell v_{BL}^2  \\
0  & 0                          &    \frac{9}{4} g_B g_\ell v_{BL}^2         &  \frac{9}{4} g_B^2 v_{BL}^2
\end{array} \right) \nonumber
\end{eqnarray}
where $v^2=v^2_1+v^2_2=\mbox{174\, GeV}^2\,$.
The complete diagonalization gives one massless photon $A$, SM neutral gauge boson
$Z$ and heavy neutral gauge boson $Z_{1,2}$.     
      

We now present a discussion of the dark matter phenomenology where we look at constraints
from direct and indirect detection experiments and collider physics.

\section{Dark Matter in LRSM}


We introduce the DM singlet fermion through the following Lagrangian term,
\begin{equation}
	\mathcal{L}_{DM} = i\bar{\chi} \gamma^\mu
	\mathcal{D}_\mu\chi-M\bar{\chi}\chi+\frac{1}{2}M_{Z_1}^2Z_{1\mu}Z^\mu_1+\frac{1}{2}M_{Z_2}^2Z_{2\mu}Z^\mu_2
	\label{ldm}
\end{equation}
where $\mathcal{D}_\mu\chi=(\partial_\mu+ig_Bn_B\,Z_{B \mu}+ig_\ell n_\ell Z_{L \mu})\chi$.
Here we denote weak gauge bosons by $Z_\ell$ and $Z_B$ while the mass eigenstates are
denoted by $Z_{1,2}$. Here $Z_B, Z_\ell$ can be written in terms of an admixture of mass
eigenstates of all neutral gauge bosons in the theory.  Assuming mixing between SM neutral
gauge boson $Z$ and other heavy gauge bosons $Z_{1,2}$ to be small ($<10^{-3}$), the
$\overline{\chi} \chi$ annihilation channels through $Z$ boson mediated processes can be
neglected. Thus, the only relevant annihilation channels for $\overline{\chi} \chi$ are
coming from $Z_{1,2}$ mediated diagrams.  The interactions of DM with SM particles which
contribute to the relic density are described by the following lagrangian term,
\begin{equation}
-g_B n^f_B \overline{f} \gamma^\mu f Z_{B \mu}
-g_\ell n^f_\ell \overline{f} \gamma^\mu f Z_{\ell \mu}\, ,
	\label{dmu}
\end{equation}
where $g_B$, $g_\ell$ are gauge coupling for $U(1)_{B,L}$ gauge groups while 
$n^f_B$, $n^f_\ell$ are baryon and lepton charges for usual fermions including quarks and
leptons.

Now, the expression for the DM relic density is given by \cite{Gondolo:1990dk}
\begin{equation}
\label{eq:relicdensity}
\Omega_\text{DM} h^2 = \frac{\unit[2.14 \times 10^{9}]{GeV}^{-1}}{J(x_f) \sqrt{g_\ast} \ M_\text{Pl}},
\end{equation}
with $J(x_f)$ written as
\begin{equation}
J(x_f)=\int_{x_f}^{\infty} \frac{ \langle \sigma v \rangle (x)}{x^2} dx.
\end{equation}

Here $\langle \sigma v\rangle$ is the thermally averaged DM annihilation cross section.
The analytical expression for the DM annihilation cross section through an intermediate
$Z_{B,\ell}$ state is given by\cite{Duerr:2015wfa}
\begin{align}
	&\sigma\left(\overline{\chi} \chi \to Z_{B,\ell}^* \to \bar{f}f \right)=
\frac{N_c^f {n_{B}^f}^2 g^4_B {n^f_{B}}^2}{12 \pi s}\nonumber\\ 
&\frac{\sqrt{s-4 m^2_f}\left(s+2 m^2_\chi \right) \left(s+2 m^2_f \right)}{\sqrt{s-4
m^2_\chi} \left(\left(s- M^2_{Z_{B}}\right)^2+ M^2_{Z_B} \Gamma^2_{Z_{B}} \right)}
\end{align}

\begin{table}[t!]
	\centering
	\begin{tabular}{|c|c|c|c|c|c|}

		\hline\hline
		Benchmark point & $M_\chi$ & $M_{Z_B}$ & $g_{B}$ &
		$n_{B}$ & $\Omega h^2$ \\\hline
		BP1	&	460  &   1000  &  0.03  &  2.0  &    0.1093\\
		BP2	&	470  &   1000  &  0.1   &  0.25 &    0.1169\\
		BP3	&	898  &   2000  &  0.1   &  2.0  &    0.1101\\
		BP4	&	925  &   2000  &  0.25  &  0.25 &    0.1104\\
		\hline\hline
		
	\end{tabular}
	\caption{Benchmark points satisfying constraints from relic density, indirect
	signals of DM and collider physics.}
	\label{tab:bmp}
\end{table}

\begin{figure}[t!]
	\centering
	\includegraphics[width=0.45\textwidth]{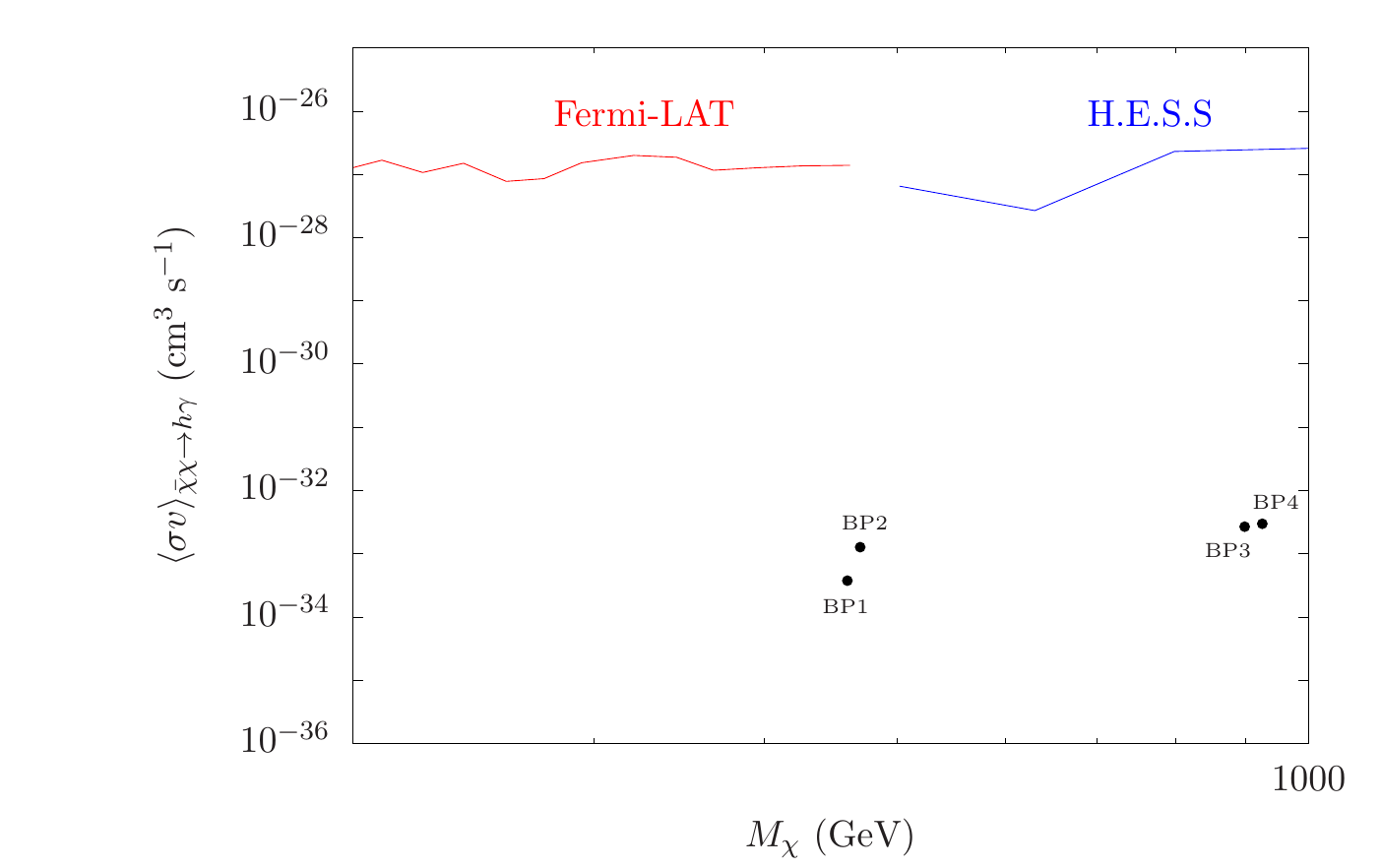}
	\includegraphics[width=0.45\textwidth]{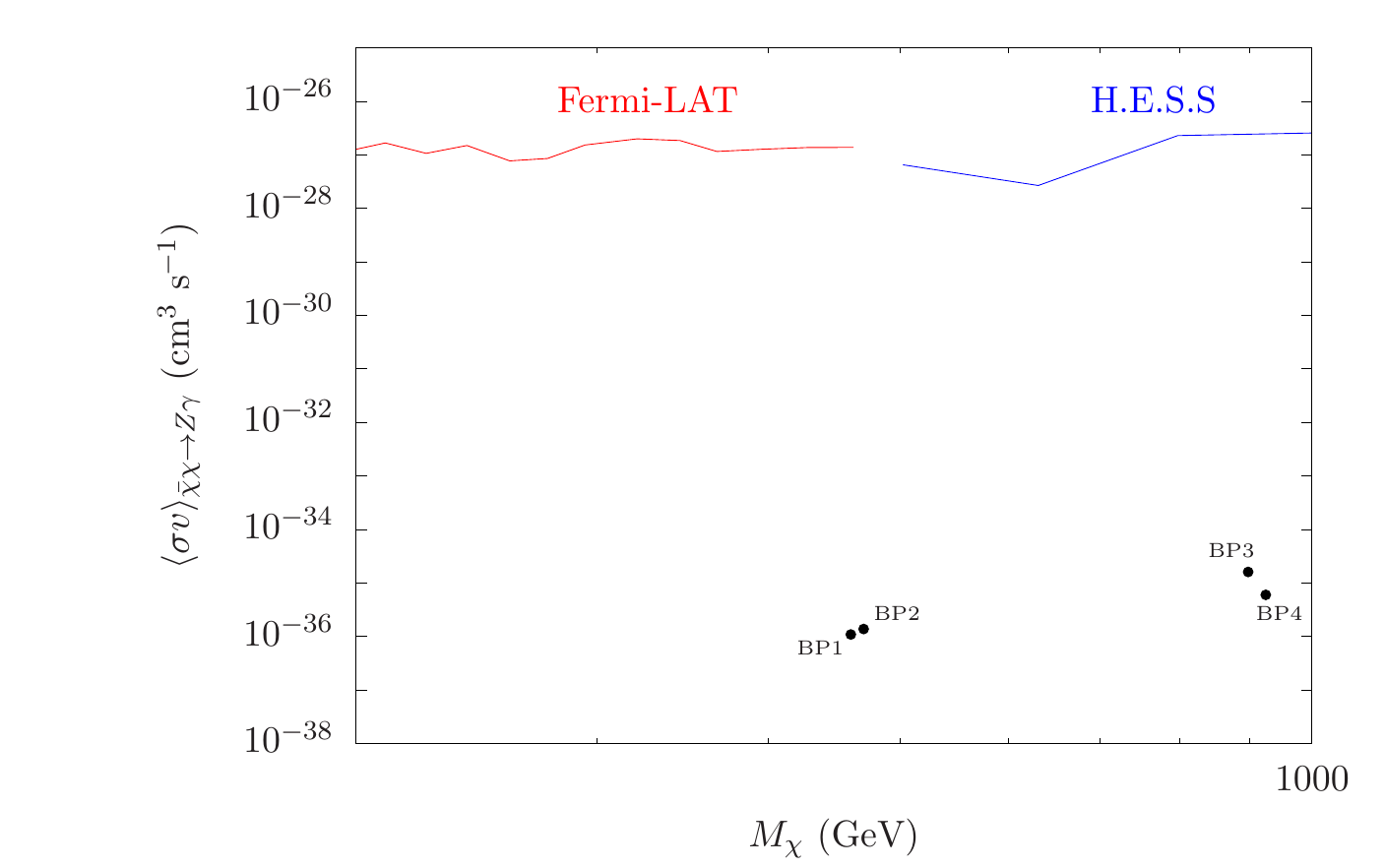}
	\caption{Gamma ray signal from the processes $\bar{\chi}\chi\to h\gamma$ (top
	panel) and $\bar{\chi}\chi\to Z\gamma$ (bottom panel) along with the observed data
	from H.E.S.S\cite{hess} and Fermi-LAT\cite{fermi-lat} for the benchmark points
	listed in Table~\ref{tab:bmp}.}
	\label{fig:gamma}
\end{figure}

\begin{figure}[t!]
	\centering
	\includegraphics[width=0.45\textwidth]{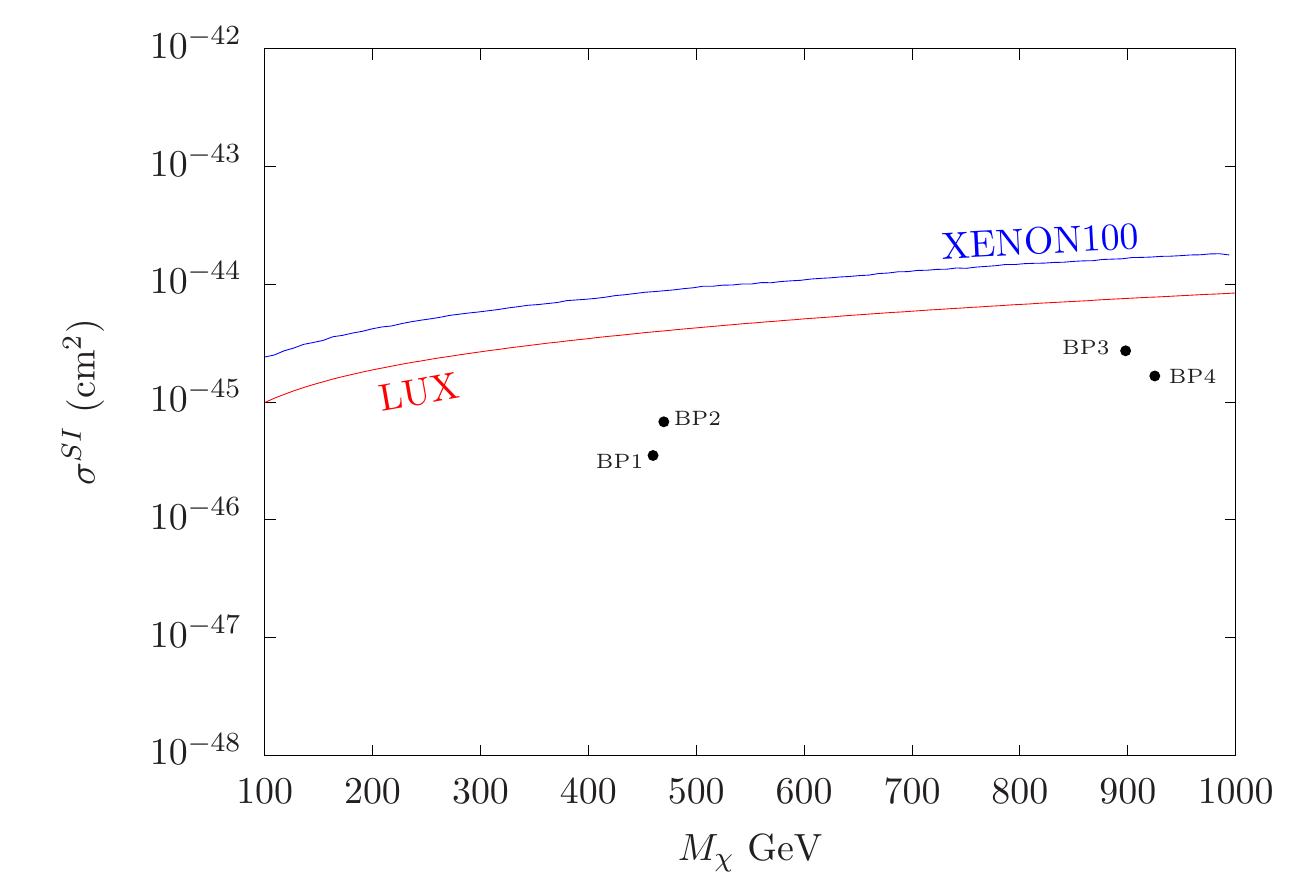}
	\caption{Spin independent cross section from DM scattering off nucleons along with
		limits from LUX\cite{lux} and XENON100\cite{xenon100} for the benchmark
		points listed in Table~\ref{tab:bmp}.}
	\label{fig:ddpl}
\end{figure}

Now in order to study the DM phenomenology we look at four specific benchmark points which
are listed in Table~\ref{tab:bmp} according to the parameters $M_\chi$, $M_{Z_B}$, $g_{B}$
and $n_{B}$.  Although the current framework includes two neutral gauge bosons
$Z_{B,\ell}$ in addition to the SM ones, we choose $Z_B$ to be the mediator for DM
interactions with SM particles while the other extra gauge boson is chosen to be super
heavy and hence does not give any observable signal.  We choose the benchmark points such
that the relic density $\Omega h^2$ lies within $5\sigma$ limit of the PLANCK value
$0.1199\pm 0.0022$\cite{Ade:2015xua}.  In each of the four benchmark points we chose
$M_{Z_B}/g_{B}\gtrsim 6$ TeV in accordance with the LEP limit on $Z^\prime$.  Here we note
that much more stringent constraints on $Z^\prime$ have recently been analysed using the
LHC data on dilepton searches which constrain $Z^\prime$ mass to be $\gtrsim 2$ TeV
\cite{Patra:2015bga,Lindner:2016lpp}.  However, these limits are for conventional cases of
LRSM which are not strictly applicable to the scenario of LRSM considered here in which
the gauge couplings corresponding to $U(1)_{B}$, $U(1)_{L}$ gauge groups are free
parameters.

\begin{figure}[t!]
	\centering
	\includegraphics[width=0.45\textwidth]{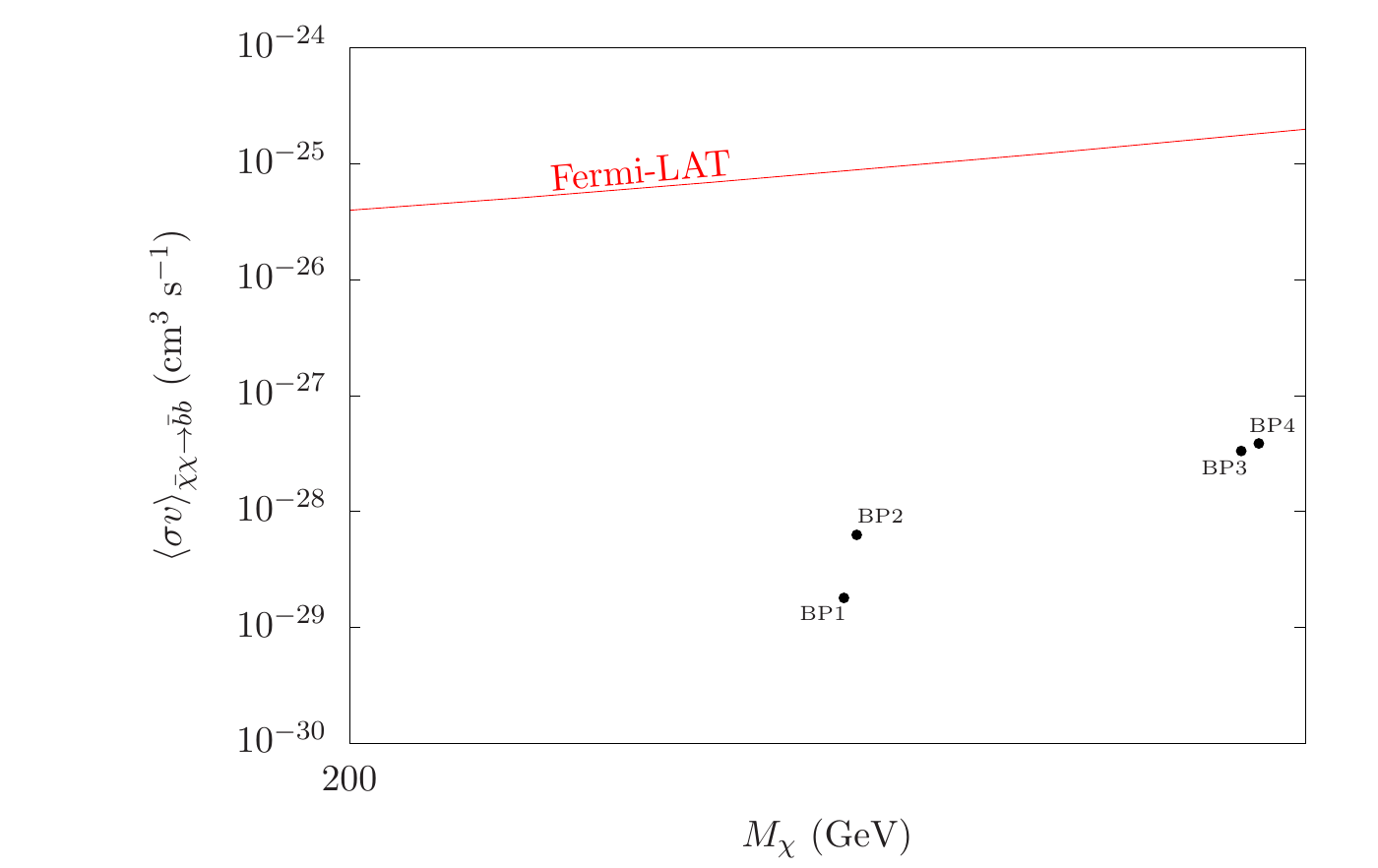}
	\caption{The DM annihilation rate for $b\bar{b}$ channel for the benchmark points
	shown in Table~\ref{tab:bmp}.  Also shown in red is the Fermi-LAT constraint from
	dwarf spheroidal galaxies on the $b\bar{b}$ annihilation rate.}
	\label{fig:bbcs}
\end{figure}

Recent limits from searches for monochromatic gamma ray emission from HESS
\cite{Abramowski:2013ax} and Fermi-LAT \cite{Ackermann:2015lka} put constraints in the
current framework of LRSM model where gamma ray line signal can be generated from the
following process \begin{align} &\bar{\chi}\chi\to Z_B^\ast \to
	\gamma\gamma,h\gamma,Z\gamma\\ &\bar{\chi}\chi\to Z_\ell^\ast \to
	\gamma\gamma,h\gamma,Z\gamma.  \label{gamma} \end{align}
In the present scenario gamma ray line signatures can be observed from processes with
final states $h\gamma$ and $Z\gamma$ while $\gamma\gamma$ is absent as it only occurs for
axial couplings of DM to $Z_B$\cite{Duerr:2015wfa}.  The gamma ray line produced in this
way could be seen distinctly in experiments since most of the astrophysical sources
produce continuum spectra.  We plot the DM signal from the $h\gamma$ and $Z\gamma$
channels along with the observed data from the gamma ray line searches of
H.E.S.S\cite{hess} and Fermi-LAT\cite{fermi-lat} as shown in Fig~\ref{fig:gamma}.  We use
the Einasto profile of DM density distribution in order to compare with the experimental
data and the analytical expressions for one loop cross section from \cite{Duerr:2015wfa}.
We see that the DM signal for the four benchmark points chosen is well below the observed
data, this is because higher values of DM coupling to $Z_B$ would conflict with the direct
detection limits whose results are shown in Fig.~\ref{fig:ddpl}.  In order to reduce the
direct detection scattering cross section the coupling of DM to $Z_B$ needs to be
decreased and this in turn results in an overabundance of relic density.  In order to
achieve the correct relic density the annihilation rate of DM needs to be increased and
this is done through increasing the mass of the $Z_B$ such that it gets closer to the
resonance value of $\sim 2M_\chi$.  In addition we also see that the constraint from dwarf
spheroidal galaxies by Fermi-LAT \cite{fermi-dwarf} is also satisfied as shown in
Fig.~\ref{fig:bbcs} where the annihilation rate for $\bar{b}b$ is plotted for the
benchmark points which satisfy relic density.  In conclusion we see that the constraints
from direct detection are the strongest on the DM mass range below $1$ TeV which we have
chosen in our benchmark points.  

\section{Conclusion}

We have shown that a singlet stable dark matter naturally arises within LRSM where baryon
and lepton numbers are promoted to separate gauge symmetries.  We point out that the
electric charge neutrality condition for DM forces it's charges under $U(1)_{B,l}$ gauge
groups to be equal. We then proceed to study the DM phenomenology in such a scenario.  We
present benchmark points that satisfy constraints from direct as well as indirect
detection experiments while being consistent with the relic density observation.  In
particular we look at constraints from gamma ray line searches by Fermi-LAT and HESS as
well as limits on $b\bar{b}$ cross section from observations on dwarf spheroidal galaxies
by Fermi-LAT and direct detection limits from LUX and XENON100.  We find that for DM mass
range below $1$ TeV the strongest constraints come from direct detection experiments.  And
finally, we note that this formalism can also be applied to singlet scalar DM which we
plan to pursue in future work.

\section{Acknowledgement} 
Both the authors acknowledge the warm hospitality provided by the organizers during the
WHEPP XIV workshop held in IIT Kanpur, India, between $4^{th} - 13^{th}$ December,  2015,
during which this work was completed. The work of SP is partially supported by the Department 
of Science and Technology, Govt.\ of India under the financial grant SB/S2/HEP-011/2013.

\bibliographystyle{utcaps_mod}
\bibliography{onubb_LR}

\providecommand{\href}[2]{#2}\begingroup\raggedright\begin{thebibliography}{10}

\bibitem{Mohapatra:1974gc}
R.~Mohapatra and J.~C. Pati, ``{\em {A Natural Left-Right Symmetry}},''
\href{http://dx.doi.org/10.1103/PhysRevD.11.2558}{Phys.Rev. {\normalfont
  \bfseries D11} (1975)  2558}.

\bibitem{Pati:1974yy}
J.~C. Pati and A.~Salam, ``{\em {Lepton Number as the Fourth Color}},''
\href{http://dx.doi.org/10.1103/PhysRevD.10.275,
  10.1103/PhysRevD.11.703.2}{Phys.Rev. {\normalfont \bfseries D10} (1974)
  275--289}.

\bibitem{Senjanovic:1975rk}
G.~Senjanovi{\'c} and R.~N. Mohapatra, ``{\em {Exact Left-Right Symmetry and
  Spontaneous Violation of Parity}},''
\href{http://dx.doi.org/10.1103/PhysRevD.12.1502}{Phys.Rev. {\normalfont
  \bfseries D12} (1975)  1502}.

\bibitem{Senjanovic:1978ev}
G.~Senjanovi{\'c}, ``{\em {Spontaneous Breakdown of Parity in a Class of Gauge
  Theories}},''
\href{http://dx.doi.org/10.1016/0550-3213(79)90604-7}{Nucl.Phys. {\normalfont
  \bfseries B153} (1979)  334--364}.

\bibitem{Mohapatra:1979ia}
R.~N. Mohapatra and G.~Senjanovi{\'c}, ``{\em {Neutrino Mass and Spontaneous
  Parity Violation}},''
\href{http://dx.doi.org/10.1103/PhysRevLett.44.912}{Phys.Rev.Lett. {\normalfont
  \bfseries 44} (1980)  912}.

\bibitem{Mohapatra:1980yp}
R.~N. Mohapatra and G.~Senjanovi{\'c}, ``{\em {Neutrino Masses and Mixings in
  Gauge Models with Spontaneous Parity Violation}},''
\href{http://dx.doi.org/10.1103/PhysRevD.23.165}{Phys.Rev. {\normalfont
  \bfseries D23} (1981)  165}.

\bibitem{Heeck:2015qra}
J.~Heeck and S.~Patra, ``{\em {Minimal Left-Right Symmetric Dark Matter}},''
  \href{http://dx.doi.org/10.1103/PhysRevLett.115.121804}{Phys. Rev. Lett.
  {\normalfont \bfseries 115} (2015) no.~12, 121804},
\href{http://arxiv.org/abs/1507.01584}{{\normalfont \ttfamily
  arXiv:1507.01584}}.

\bibitem{Bezrukov:2009th}
F.~Bezrukov, H.~Hettmansperger, and M.~Lindner, ``{\em {keV sterile neutrino
  Dark Matter in gauge extensions of the Standard Model}},''
  \href{http://dx.doi.org/10.1103/PhysRevD.81.085032}{Phys. Rev. {\normalfont
  \bfseries D81} (2010)  085032},
\href{http://arxiv.org/abs/0912.4415}{{\normalfont \ttfamily arXiv:0912.4415}}.

\bibitem{Nemevsek:2012cd}
M.~Nemevsek, G.~Senjanovic, and Y.~Zhang, ``{\em {Warm Dark Matter in Low Scale
  Left-Right Theory}},''
  \href{http://dx.doi.org/10.1088/1475-7516/2012/07/006}{JCAP {\normalfont
  \bfseries 1207} (2012)  006},
\href{http://arxiv.org/abs/1205.0844}{{\normalfont \ttfamily arXiv:1205.0844}}.

\bibitem{Barry:2014ika}
J.~Barry, J.~Heeck, and W.~Rodejohann, ``{\em {Sterile neutrinos and
  right-handed currents in KATRIN}},''
  \href{http://dx.doi.org/10.1007/JHEP07(2014)081}{JHEP {\normalfont \bfseries
  07} (2014)  081},
\href{http://arxiv.org/abs/1404.5955}{{\normalfont \ttfamily arXiv:1404.5955}}.

\bibitem{Garcia-Cely:2015quu}
C.~Garcia-Cely and J.~Heeck, ``{\em {Phenomenology of left-right symmetric dark
  matter}},'' \href{http://arxiv.org/abs/1512.03332}{{\normalfont \ttfamily
  arXiv:1512.03332}}.
[JCAP1603,021(2016)].

\bibitem{Ade:2015xua}
{\normalfont \bfseries Planck}, P.~A.~R. Ade {\em et al.}, ``{\em {Planck 2015
  results. XIII. Cosmological parameters}},''
\href{http://arxiv.org/abs/1502.01589}{{\normalfont \ttfamily
  arXiv:1502.01589}}.

\bibitem{Duerr:2013dza}
M.~Duerr, P.~Fileviez~Perez, and M.~B. Wise, ``{\em {Gauge Theory for Baryon
  and Lepton Numbers with Leptoquarks}},''
  \href{http://dx.doi.org/10.1103/PhysRevLett.110.231801}{Phys. Rev. Lett.
  {\normalfont \bfseries 110} (2013)  231801},
\href{http://arxiv.org/abs/1304.0576}{{\normalfont \ttfamily arXiv:1304.0576}}.

\bibitem{He:1989mi}
X.-G. He and S.~Rajpoot, ``{\em {Anomaly Free Left-right Symmetric Models With
  Gauged Baryon and Lepton Numbers}},''
\href{http://dx.doi.org/10.1103/PhysRevD.41.1636}{Phys. Rev. {\normalfont
  \bfseries D41} (1990)  1636}.

\bibitem{Duerr:2013opa}
M.~Duerr, P.~Fileviez~Perez, and M.~Lindner, ``{\em {Left-Right Symmetric
  Theory with Light Sterile Neutrinos}},''
  \href{http://dx.doi.org/10.1103/PhysRevD.88.051701}{Phys. Rev. {\normalfont
  \bfseries D88} (2013)  051701},
\href{http://arxiv.org/abs/1306.0568}{{\normalfont \ttfamily arXiv:1306.0568}}.

\bibitem{Gondolo:1990dk}
P.~Gondolo and G.~Gelmini, ``{\em {Cosmic abundances of stable particles:
  Improved analysis}},''
\href{http://dx.doi.org/10.1016/0550-3213(91)90438-4}{Nucl. Phys. {\normalfont
  \bfseries B360} (1991)  145--179}.

\bibitem{Duerr:2015wfa}
M.~Duerr, P.~Fileviez~Perez, and J.~Smirnov, ``{\em {Simplified Dirac Dark
  Matter Models and Gamma-Ray Lines}},''
  \href{http://dx.doi.org/10.1103/PhysRevD.92.083521}{Phys. Rev. {\normalfont
  \bfseries D92} (2015) no.~8, 083521},
\href{http://arxiv.org/abs/1506.05107}{{\normalfont \ttfamily
  arXiv:1506.05107}}.

\bibitem{hess}
{\normalfont \bfseries HESS}, A.~Abramowski {\em et al.}, ``{\em {Search for
  Photon-Linelike Signatures from Dark Matter Annihilations with H.E.S.S.}},''
  \href{http://dx.doi.org/10.1103/PhysRevLett.110.041301}{Phys. Rev. Lett.
  {\normalfont \bfseries 110} (2013)  041301},
\href{http://arxiv.org/abs/1301.1173}{{\normalfont \ttfamily arXiv:1301.1173}}.

\bibitem{fermi-lat}
{\normalfont \bfseries Fermi-LAT}, M.~Ackermann {\em et al.}, ``{\em {Updated
  search for spectral lines from Galactic dark matter interactions with pass 8
  data from the Fermi Large Area Telescope}},''
  \href{http://dx.doi.org/10.1103/PhysRevD.91.122002}{Phys. Rev. {\normalfont
  \bfseries D91} (2015) no.~12, 122002},
\href{http://arxiv.org/abs/1506.00013}{{\normalfont \ttfamily
  arXiv:1506.00013}}.

\bibitem{lux}
{\normalfont \bfseries LUX}, D.~S. Akerib {\em et al.}, ``{\em {First results
  from the LUX dark matter experiment at the Sanford Underground Research
  Facility}},'' \href{http://dx.doi.org/10.1103/PhysRevLett.112.091303}{Phys.
  Rev. Lett. {\normalfont \bfseries 112} (2014)  091303},
\href{http://arxiv.org/abs/1310.8214}{{\normalfont \ttfamily arXiv:1310.8214}}.

\bibitem{xenon100}
{\normalfont \bfseries XENON100}, E.~Aprile {\em et al.}, ``{\em {Dark Matter
  Results from 225 Live Days of XENON100 Data}},''
  \href{http://dx.doi.org/10.1103/PhysRevLett.109.181301}{Phys. Rev. Lett.
  {\normalfont \bfseries 109} (2012)  181301},
\href{http://arxiv.org/abs/1207.5988}{{\normalfont \ttfamily arXiv:1207.5988}}.

\bibitem{Patra:2015bga}
S.~Patra, F.~S. Queiroz, and W.~Rodejohann, ``{\em {Stringent Dilepton Bounds
  on Left-Right Models using LHC data}},''
  \href{http://dx.doi.org/10.1016/j.physletb.2015.11.009}{Phys. Lett.
  {\normalfont \bfseries B752} (2016)  186--190},
\href{http://arxiv.org/abs/1506.03456}{{\normalfont \ttfamily
  arXiv:1506.03456}}.

\bibitem{Lindner:2016lpp}
M.~Lindner, F.~S. Queiroz, and W.~Rodejohann, ``{\em {Dilepton bounds on
  left-right symmetry at the LHC run II and neutrinoless double beta decay}},''
\href{http://arxiv.org/abs/1604.07419}{{\normalfont \ttfamily
  arXiv:1604.07419}}.

\bibitem{Abramowski:2013ax}
{\normalfont \bfseries HESS}, A.~Abramowski {\em et al.}, ``{\em {Search for
  Photon-Linelike Signatures from Dark Matter Annihilations with H.E.S.S.}},''
  \href{http://dx.doi.org/10.1103/PhysRevLett.110.041301}{Phys. Rev. Lett.
  {\normalfont \bfseries 110} (2013)  041301},
\href{http://arxiv.org/abs/1301.1173}{{\normalfont \ttfamily arXiv:1301.1173}}.

\bibitem{Ackermann:2015lka}
{\normalfont \bfseries Fermi-LAT}, M.~Ackermann {\em et al.}, ``{\em {Updated
  search for spectral lines from Galactic dark matter interactions with pass 8
  data from the Fermi Large Area Telescope}},''
  \href{http://dx.doi.org/10.1103/PhysRevD.91.122002}{Phys. Rev. {\normalfont
  \bfseries D91} (2015) no.~12, 122002},
\href{http://arxiv.org/abs/1506.00013}{{\normalfont \ttfamily
  arXiv:1506.00013}}.

\bibitem{fermi-dwarf}
{\normalfont \bfseries Fermi-LAT}, M.~Ackermann {\em et al.}, ``{\em {Searching
  for Dark Matter Annihilation from Milky Way Dwarf Spheroidal Galaxies with
  Six Years of Fermi Large Area Telescope Data}},''
  \href{http://dx.doi.org/10.1103/PhysRevLett.115.231301}{Phys. Rev. Lett.
  {\normalfont \bfseries 115} (2015) no.~23, 231301},
\href{http://arxiv.org/abs/1503.02641}{{\normalfont \ttfamily
  arXiv:1503.02641}}.

\end{thebibliography}\endgroup
\end{document}